\documentclass[preprint,pre,amsmath,amssymb,superscriptaddress]{revtex4-1}

\usepackage{graphicx}
\usepackage{bm}
\usepackage{color}
\usepackage{dcolumn}
\usepackage{ulem}
\usepackage{hhline}
\usepackage{float}

\begin{document}

\title{Statistics of adatom diffusion in a model of thin film growth}

\author{Edwin E. Mozo Luis}
\email{eluis@ufba.br}
\affiliation{Instituto de F\'{\i}sica, Universidade Federal da Bahia,
  Campus Universit\'{a}rio da Federa\c c\~ao,
  Rua Bar\~{a}o de Jeremoabo s/n, 40170-115, Salvador, BA, Brazil}
\author{Ismael S. S. Carrasco}
\email{theismiu@gmail.com}
\affiliation{Instituto de F\'{i}sica, Universidade Federal Fluminense,
Avenida Litor\^{a}nea s/n, 24210-340 Niter\'{o}i, RJ, Brazil}
\author{Thiago A. de Assis}
\email{thiagoaa@ufba.br}
\affiliation{Instituto de F\'{\i}sica, Universidade Federal da Bahia,
  Campus Universit\'{a}rio da Federa\c c\~ao,
  Rua Bar\~{a}o de Jeremoabo s/n, 40170-115, Salvador, BA, Brazil}
\author{F\'abio D. A. Aar\~ao Reis}
\email{reis@if.uff.br}
\affiliation{Instituto de F\'{i}sica, Universidade Federal Fluminense,
Avenida Litor\^{a}nea s/n, 24210-340 Niter\'{o}i, RJ, Brazil}

\begin{abstract}

We study the statistics of the number of executed hops of adatoms at the surface of films
grown with the Clarke-Vvedensky (CV) model in simple cubic lattices.
The distributions of this number, $N$, are determined in
films with average thicknesses close to $50$ and $100$ monolayers for a broad
range of values of the diffusion-to-deposition ratio $R$
and of the probability $\epsilon$ that lowers the diffusion coefficient for each lateral
neighbor.
The mobility of subsurface atoms and the energy barriers for crossing step edges
are neglected.
Simulations show that the adatoms execute uncorrelated diffusion during the
time in which they move on the film surface.
In a low temperature regime, typically with $R\epsilon\lesssim 1$, the attachment to lateral
neighbors is almost irreversible, the average number of hops scales as
$\langle N\rangle \sim R^{0.38\pm 0.01}$, and the distribution of that number
decays approximately as $\exp\left[-\left({N/\langle N\rangle}\right)^{0.80\pm 0.07}\right]$.
Similar decay is observed in simulations of random walks in a plane with randomly
distributed absorbing traps and the estimated relation between $\langle N\rangle$ and the
density of terrace steps is similar to that observed in the trapping problem,
which provides a conceptual explanation of that regime.
As the temperature increases, $\langle N\rangle$ crosses over to another regime when
$R\epsilon^{3.0\pm 0.3}\sim 1$, which indicates high mobility of
all adatoms at terrace borders.
The distributions $P\left( N\right)$ change to simple exponential decays, due to the constant
probability for an adatom to become immobile after being covered by a new deposited layer.
At higher temperatures, the surfaces become very smooth and
$\langle N\rangle \sim R\epsilon^{1.85\pm 0.15}$,
which is explained by an analogy with submonolayer growth.
Thus, the statistics of adatom hops on growing film surfaces
is related to universal and non-universal features of the growth model and with properties
of trapping models if the hopping time is limited by the landscape
and not by the deposition of other layers.

\end{abstract}

\maketitle

\section{Introduction}
\label{intro}

Modeling thin film deposition has been a topic of interest for several decades due to
the large number of technological applications of those materials \citep{ohring,michely}
and the connection with non-equilibrium Statistical Mechanics \cite{barabasi,krug}.
For the deposition of samples with large crystalline grains, an essential
ingredient is the surface diffusion of adsorbed atoms (adatoms) or molecules, which favors
the aggregation at low energy sites.
For this reason, the models have to represent the interplay between the atomic flux and the
adatom diffusion, the latter being described by hops between sites of a crystal surface.
Using kinetic Monte Carlo (kMC) simulations and analytical methods,
such collective diffusion models have already reproduced morphological features of several materials
and provided estimates of energy barriers for microscopic processes
(adsorption, diffusion, and desorption) \citep{pimpinelli,etb}.

The simplest model of this type is that of Clarke and Vvedensky (CV) \cite{cv1,cv},
in which the temperature activated diffusion of an adatom includes a terrace contribution
and a term additive over the lateral neighbors; in this version,
adsorption barriers and barriers for crossing step edges are neglected.
The CV model was already used to determine universal properties in
submonolayer and multilayer growth \citep{ratsch1995,bartelt1995,etb,submonorev},
was studied in the context of kinetic roughening
\citep{kotrla1996,dassarma1996,haselwandter2007,haselwandterPRE2010,cv2015,mozo2017,martynec2019},
and extended to molecular and colloidal particle film deposition
\citep{ganapathy,bommel2014,kleppmann2017}.

As a film grows, the surface diffusion of an adatom occurs in a certain time interval
between its adsorption and its attachment at a final position of the crystal.
This process may be followed by subsurface or bulk diffusion,
but this feature is neglected in most deposition models so that the film morphology
is solely determined by the surface dynamics.
The possible fundamental and applied interest on the statistics of the surface diffusion of
individual adatoms motivates the present work, in which
the distributions of the number $N$ of hops executed by the surface adatoms is
studied in the CV model.
From a theoretical point of view, this type of investigation may help
the description of the film morphology in this widely studied model and may be useful
for related growth models.
Moreover, with the advance in microscopy techniques, particularly in scanning tunneling
microscopy, it is possible to monitor the movement of individual atoms and molecules
\citep{Besenbacher,Besenbacher2,Golovchenko,
Baro,mayne2001,bussmann2008,bussmann2010,
tringides2010,mielke2016}, so the statistics of adatom diffusion lengths and diffusion times
may be accessed.

Our numerical study of the distribution $P\left( N\right)$ and of
the average number of executed hops $\langle N\rangle$ distinguishes two scaling regimes.
These regimes correspond, respectively, to conditions in which aggregation to lateral neighbors
is almost irreversible (called low temperature regime) and in which the detachment of all adatoms
from terrace edges is facile (called high temperature regime).
The scaling properties in the low temperature regime are explained by a connection with
the problem of random walkers in a plane with static absorbing traps
\citep{donsker1979,bunde1997,havlin}, whose effective density in the film surface
can be related to non-universal (temperature-dependent) properties of the height fluctuations
\cite{CDLM}.
The high temperature regime is characterized by the smoothness of the film surfaces and
the scaling properties have relations with those of submonolayer deposition
\citep{ratsch1995,bartelt1995,submonorev}.
These results show that a combination of kinetic roughening concepts with
scaling properties of trapping processes may be used to
understand the statistics of the adatom hops in growing films,
particularly when their diffusion is limited by the film landscape and not by the deposition.
This may eventually help to determine surface diffusion lengths in more realistic models,
which can be compared with experimental results.

The rest of this paper is organized as follows.
In Section \ref{basics}, we present the CV model, the simulation methods, and
the quantities to be measured.
We also introduce a limited mobility (LM) model with similar roughening properties,
which helps the interpretation of the numerical results.
In Section \ref{results}, we show the results of kMC simulations for the average diffusion
length and for its distribution, including the derivation of a scaling relation to connect
low and high temperature data.
The results for the LM model are also presented.
In Section \ref{discussion}, we use scaling approaches and results of other models to
explain universal and non-univeral scaling properties of $P\left( N\right)$.
In Section \ref{conclusion}, a summary of results and conclusions is presented.

\section{Model and methods}
\label{basics}

\subsection{The CV model}
\label{CVmodel}

The CV model is defined in a simple cubic lattice in which the edge of a site is
the unit length.
The initially flat substrate is located at $z=0$, with lateral size $L$ and
periodic boundary conditions in the $x$ and $y$ directions.
Solid-on-solid conditions are considered, so no overhangs are allowed at the film surface.
A column of the deposit is defined as the set of adatoms with the same $\left( x,y\right)$
position;
the height variable $h\left( x,y\right)$ is the maximal height of an adatom in that column.

The deposition occurs with a collimated flux of $F$ atoms per substrate site per unit time.
In each deposition event, a column $\left( x,y\right)$ is randomly chosen and the new atom
is adsorbed as it lands at the top of that column.

\begin{figure}[!h]
\center
\includegraphics [scale=0.3]{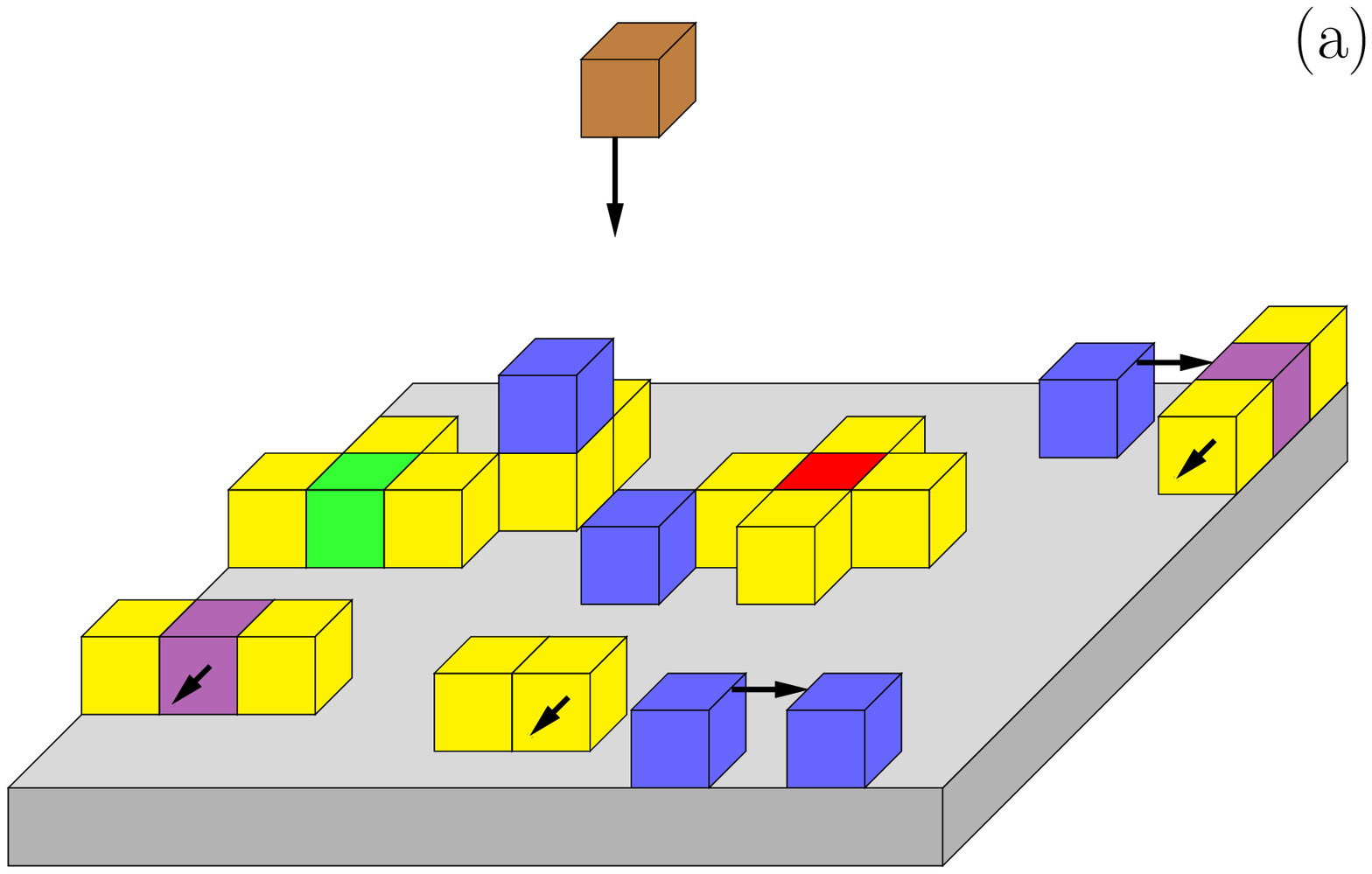} \hskip 0.5cm
\includegraphics [scale=0.3]{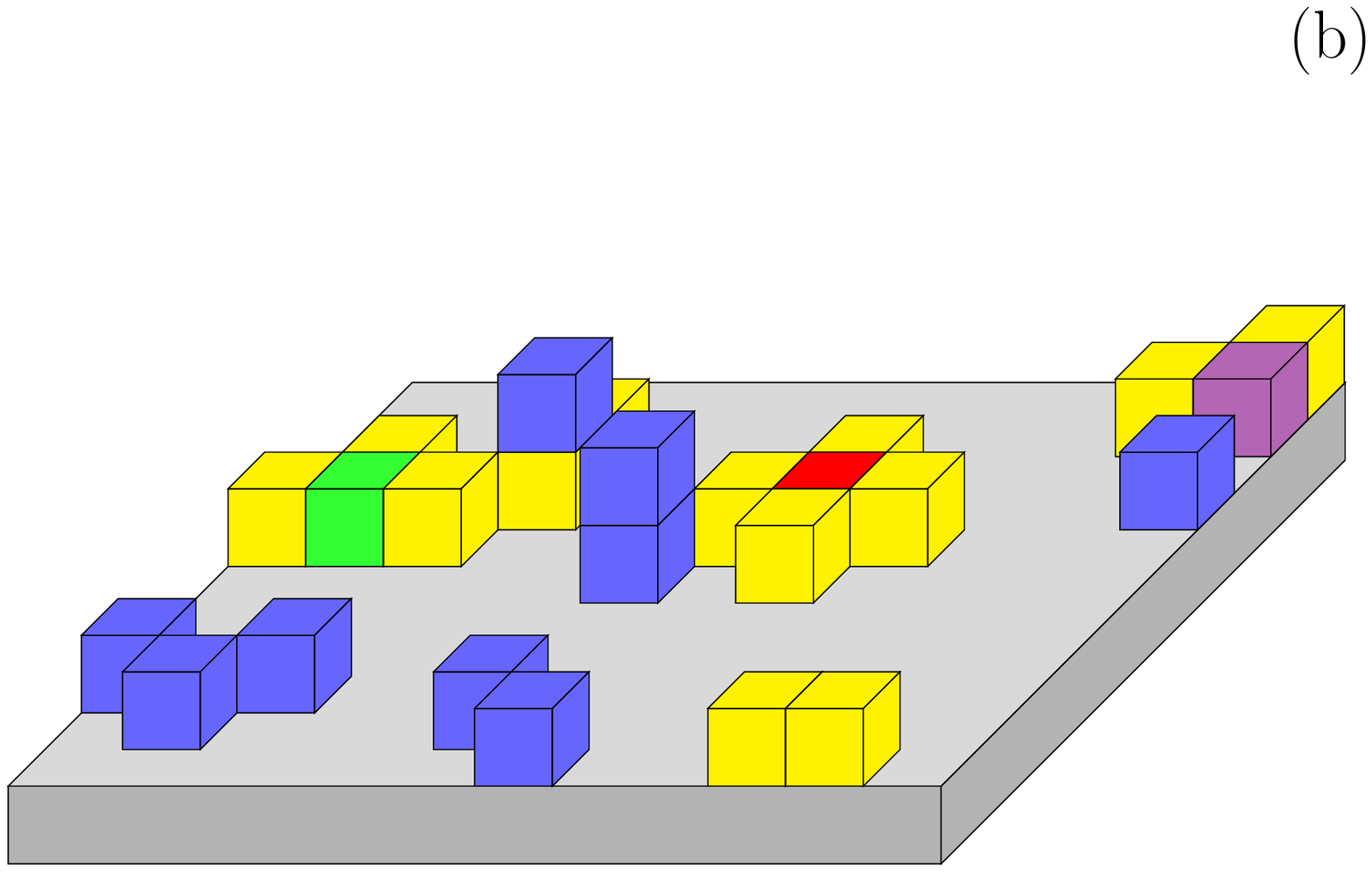}
\caption{Illustration adatom hops considering a small section of a system with $L=512$:
(a) the incidence of a new atom and the directions of hops of five adatoms are indicated by arrows;
(b) the configuration after the adsorption of the new atom and the five hops.
The inert substrate is shown in gray and
the color of each adatom indicates the current number of lateral neighbors: $n=0$, blue; $n=1$, yellow;
$n=2$, magenta; $n=3$, green; $n=4$, red.}
\label{hops}
\end{figure}

Surface diffusion is simultaneously modeled by hops of the adatoms at the top of the $L^2$ columns,
with rates that depend on their local neighborhoods.
Only this set of adatoms is assumed to be mobile.
The hopping rate of an adatom in the middle of a terrace, where it has no lateral
nearest neighbor (NN), is
\begin{equation}
D_0=\nu\exp{\left( -E_s/k_BT\right)}
\label{defD0}
\end{equation}
where $\nu$ is a frequency, $E_s$ is an activation energy, and $T$ is the temperature.
If an adatom has $n$ lateral NNs, its hopping rate is
\begin{equation}
D=D_0\epsilon^n \qquad , \qquad \epsilon \equiv \exp{\left( -E_b/k_BT\right)} ,
\label{defD}
\end{equation}
where $E_b$ is the absolute value of a bond energy.
Thus, $\epsilon$ may be interpreted as a detachment probability per lateral neighbor.
The direction of each hop is randomly chosen among the four nearest neighbor (NN) columns
($\pm x$, $\pm y$) and the adatom moves to the top of that column, independently of
the height differences.
Fig. \ref{hops} illustrates the possible hops of some adatoms.

During the growth, the number of deposited layers is defined as the average number of deposited
atoms per substrate site and equals $Ft$, where $t$ is the growth time.
It corresponds to an effective deposited mass,
but does not mean that the film grows layer by layer due to the fluctuations in the flux,
even in the presence of relaxation by diffusion.

In the original CV model \citep{cv}, $\nu=2k_BT/h$, where $h$ is the Planck's constant,
as predicted by transition state theory.
However, it is more frequent that a constant value $\nu= {10}^{12}{\text s}^{-1}$ is used
in simulation and analytical works \citep{etb}, and so this value is used here.
The values of $E_s$ and $E_b$ are determined by material properties, while
$T$ and $F$ depend on the deposition conditions.
A diffusion-to-deposition ratio, which is defined as
\begin{equation}
R \equiv \frac{D_0}{F} = \frac{\nu}{F} \exp{\left( -E_s/k_BT\right)} ,
\label{defR}
\end{equation}
is the most important quantity to describe the interplay between temperature and flux.
It may be interpreted as the average number of hops of an adatom on a terrace during the
average time $1/F$ of deposition of one atomic layer.
In the study of scaling properties of the model, $R$ and $\epsilon$ are taken as the
independent parameters.

This version of the CV model neglects energy barriers for adatoms to cross step edges
(which significantly affect the morphology of real films).
Although the effects of step edge barriers make the simulations more realistic, the main reason for neglecting them here is to avoid considering details that are better introduced in system-specific scenarios.
Here we also consider the same hopping rates on the substrate and on other layers, which is
suitable to model homoepitaxial growth.
In heteroepitaxial growth, if the energetics on the substrate and in other layers is
not very different, similar results may be obtained after the deposition of several layers.

\subsection{Simulation parameters and quantities of interest}
\label{calculation}

Simulations are performed on lattices with $L=512$ considering $F={10}^{-2}{\text s}^{-1}$,
i.e. a flux of $0.01$ monolayers per second, which gives $\nu/F = {10}^{14}$.
The values of $E_s$, $E_b$, and the temperature ranges of the parameter sets labeled from
A to I are listed in Table \ref{energias}, with the corresponding ranges of $R$ and $\epsilon$.
For each parameter set, $50$ different deposits were grown.

\begin{table*}[t]
\centering
\begin{tabular}{|c|c|c|c|c|c|}
    \hline
Sets    & $E_{s}$ (eV) &$E_{b}$ (eV)  & $T$ ($K$)  & R &$\epsilon$  \\
    \hline
 A   & 0.2  &0.2  &  $[115,144]$ & $[1.7\times10^{5},1.0\times10^{7}]$ & $[1.7\times10^{-9},1.0\times10^{-7}]$ \\
    \hline
 B   &0.25  &0.2  & $[139,180]$ &$[8.6\times10^{4},1.00\times10^{7}]$  & $[5.6\times10^{-8},2.51\times10^{-6}]$  \\
    \hline
 C   &0.3  &0.2  &$[160,216]$  & $[3.6\times10^{4},1.0\times10^{7}]$  & $[5.0\times10^{-7},2.2\times10^{-5}]$  \\
    \hline
 D  & 0.4 &0.2  &$[225,289]$  & $[1.1\times10^{5},1.1\times10^{7}]$ & $[3.3\times10^{-5},3.3\times10^{-4}]$  \\
    \hline
 E   &0.4  &0.05  &$[190,270]$  &$[2.5\times10^{3},3.4\times10^{6}]$  &$[4.7\times10^{-2},0.11]$  \\
    \hline
 F   & 0.4 &0.08  &$[204,253]$  & $[1.3\times10^{4},1.1\times10^{6}]$ & $[1.1\times10^{-2},2.5\times10^{-2}]$  \\
    \hline
 G   &0.5  &0.2  &$[272,345]$  &$[5.4\times10^{4},5.0\times10^{6}]$  & $[2.0\times10^{-4},1.2\times10^{-3}]$ \\
    \hline
 H   & 0.6 &0.11  &$[300,480]$  &$[8.3\times10^{3},5.0\times10^{7}]$  &$[1.4\times10^{-2},7.0\times10^{-2}]$  \\
    \hline
 I   & 1 &0.3  &$[560,720]$  & $[1.0\times10^{5},1.0\times10^{7}]$ &$[2.0\times10^{-3},7.9\times10^{-3}]$  \\
    \hline

\end{tabular}
\caption{Values of the activation and bond energies, temperature ranges, and the corresponding
ranges of $R$ and $\epsilon$ considered in this work.}
\label{energias}
\end{table*}

If the ratio $E_s/E_b$ is constant, the same pairs of
parameters $\left( R,\epsilon\right)$ can be reached by varying the temperature.
For this reason, the values of $E_s$ in Table \ref{energias} were systematically changed
between $0.2$eV and $1.0$eV, but this was not the case for $E_b$.
The temperature ranges considered here are limited for two reasons:
first, we restrict our analysis to cases with average number of adatom hops
near $10$ or larger, i.e. cases in which the adatoms do not have their average motion
restricted to a close neighborhood of the incidence point;
second, we ensure that the average diffusion length is much smaller than the lateral size $L$,
which avoids finite size effects.

For each parameter set, we measure the number of hops $N$ of each adatom deposited
in two time intervals, $45\leq Ft\leq 50$ and $95\leq Ft\leq 100$, which are hereafter denoted
as T50 and T100, respectively.
The study in different time intervals is important to check for possible effects of the
deposition time on the statistics of the number $N$.

For each atom deposited in T50, we compute the total number of hops $N$ executed between its
adsorption and a final monitoring time $Ft\leq 65$; at this time, an atom deposited up to
$Ft\leq 50$ is certainly buried by other layers and cannot execute surface diffusion anymore,
so monitoring at longer times is not necessary.
This atom may have stopped moving after attachment to lateral neighbors or in the middle
of a terrace, but these different possibilities are not distinguished in our statistics.
The statistics in the interval T100 is obtained by monitoring the motion of the corresponding
adatoms until $Ft\leq 115$.
With this method, each time interval gives $\sim 6.5\times {10}^7$ estimates of $N$, so
we obtain accurate distributions $P\left( N\right)$ and accurate averages.

In the intervals T50 and T100, we also calculated the surface roughness, defined as the rms
fluctuation of the height distribution:
\begin{equation}
W \equiv {\left< {\left[ \overline{{\left( h - \overline{h}\right) }^2}  \right] } \right>}^{1/2}  ,
\label{defw}
\end{equation}
where the overbars indicate spatial averages
and the angular brackets indicate configurational averages.
The roughness is useful to discuss the mechanisms involved in the scaling of
$P\left( N\right)$, particularly at high temperatures.

We calculated $P\left( N\right)$ only in the intervals T50 and T100 because the algorithm for
growing films in large substrates is very time consuming for $R\gtrsim {10}^6$.
If we extended the simulations to longer times, we would have a proportionally smaller number
of configurations and the accuracy of the average quantities would be poorer.
Moreover, measurements at shorter times were not performed to avoid transient effects
in the early stages of multilayer growth.

\subsection{Simulation method}
\label{simulationmethod}

The simulations are implemented with the algorithm detailed described in Ref. \protect\cite{lam1997},
which we have also used in previous works \citep{cv2015,tosousareis2018,toreis2020}.

The $L^2$ surface atoms have their positions ($x,y$) grouped into five lists $X_{n}$
($n=0, \dots, 4$) according to the number $n$ of atoms at NN sites at the same height,
i.e. lateral neighbors.
The position of a surface atom in a list $X_n$ is stored in an inverted-list matrix
$M_1(x,y)$.
In addition, a matrix $H(x,y)$ stores the column heights, and is used for rapid access to
the configuration of the neighborhood of a mobile atom.

At each step of the simulation, the rates of all possible events (namely, deposition of a new
atom and hop of one of the $L^2$ surface atoms) are calculated and their sum is
denoted as $\Sigma$.
The probability of each event is the ratio between its rate and $\Sigma$.
Since all atoms in a list $X_n$ have the same hopping rate, the probability of that
list is the product of the number of atoms in the list and the hopping rate divided by $\Sigma$.
The event to be executed is then chosen according to those probabilities.
In the case of choosing a list, one of its atoms is randomly chosen to hop, in a direction
which is also randomly chosen among four possibilities ($\pm x$, $\pm y$).
After this simulation step, the time is incremented by $1/\Sigma$ minus the natural logarithm of a
randomly chosen number in the interval $\left( 0,1\right]$; the latter contribution has a very small
effect on the total deposition time.

\subsection{The limited mobility model}
\label{LMmodel}

The LM model studied here is an extension of the models proposed in
Ref. \protect\cite{tosousareis2018} to approximate the CV model and
in Refs. \protect\cite{edcross,capriotalebreis2018} to simulate electrodeposition.
In this type of model, the surface diffusion of each adatom is executed before the next
atom is deposited.
The lattice geometry is the same described in Sec. \ref{CVmodel}.

The adsorption of each atom occurs at the top of a randomly chosen column.
The adatom diffusion is represented as follows:

\par\noindent (i) After the incidence, it executes $G$ attempts to hop to
randomly chosen NN columns.
The probability of performing the hop is $P_{hop}=P^n$,
where $n$ is the number of lateral neighbors; with probability $1-P_{hop}$, the hop attempt
is rejected.
Thus, $P$ is a probability of detachment per lateral NN.

\par\noindent (ii) If the adatom detaches from one or more lateral NNs before executing the $G$
attempts, then it is allowed to execute a new sequence of $G$ hop attempts,
following the same rules of (i).
In other words, the counter of the number of hop attempts is reset after each detachment from
lateral neighbors.

\par\noindent (iii) If the adatom executed $G$ hop attempts and the condition in (ii) was not
satisfied, then it permanently aggregates at the current position.
This may happen when the adatom is in the middle of a terrace or when it has lateral NNs.

We performed simulations in lattices with $L=1024$
for several values of $G$ between $10$ and $60$, and for $P=0.01$ and $0.1$.
For each parameter set, $100$ deposits with a maximal average thickness of $100$ layers
were grown.
The distributions $P_{LM}\left( N\right)$ of the numbers of executed hops were obtained
in the time intervals T50 and T100, as defined in
Sec. \ref{calculation}.
An adatom may execute an arbitrarily large number of hops
due to the resetting of the number of attempts.

\section{Numerical results}
\label{results}

\subsection{Confirmation of normal adatom diffusion}
\label{diffusion}

In some of our simulations, we measured the square displacement in the horizontal directions,
$r_H^2\equiv {\left( \Delta x\right)}^2 + {\left( \Delta y\right)}^2$,
of each adatom in the interval T100.
For each $N$, we calculated the mean square displacement ${\langle r_H^2\rangle}_N$,
i.e. the mean square displacement of the adatoms that executed exactly $N$ hops.
${\langle r_H^2\rangle}_N$ is the square diffusion length of this set of adatoms
measured in the substrate directions.

\begin{figure}[!h]
\center
\includegraphics [scale=0.85]{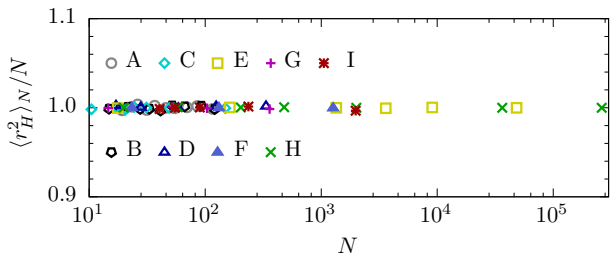}
\caption{Ratio ${\langle r_H^2\rangle}_N /N$ as a function of $N$ for
the sets of parameters, as shown in Table \ref{energias}.}
\label{normaldif}
\end{figure}

Fig. \ref{normaldif} shows the ratio ${\langle r_H^2\rangle}_N /N$ as a function of $N$ for
several parameter sets and several temperatures in each of them.
For small or large values of $N$, that ratio is very close to $1$ with an accuracy better
than $0.02\%$,
which is consistent with the absence of correlations in subsequent adatom hops.
This is expected from the CV model rules because the energy barriers for the hops depend on
the local neighborhoods but not on the directions of the hop attempts.

Due to the normal diffusion, the distributions of displacements as a function of $N$ are Gaussian
when $N\gg 1$.
They may be obtained in terms of the model parameters using the
the distributions of the number of hops $N$ calculated below.

\subsection{Average number of hops at low temperatures}
\label{averagelow}

This is typically the case of $R\epsilon \ll 1$, in which an adatom
with one or more lateral NNs has a small probability to move during the time $1/F$ of
deposition of a new atomic layer.
In these conditions, the attachment of the adatom to lateral NNs is expected to be almost
irreversible.
The average number of hops $\langle N\rangle$ then depends only on the parameter $R$,
similarly to the case $\epsilon=0$; see Ref. \protect\cite{CDLM}.

\begin{figure}[!h]
\center
\includegraphics [scale=0.9]{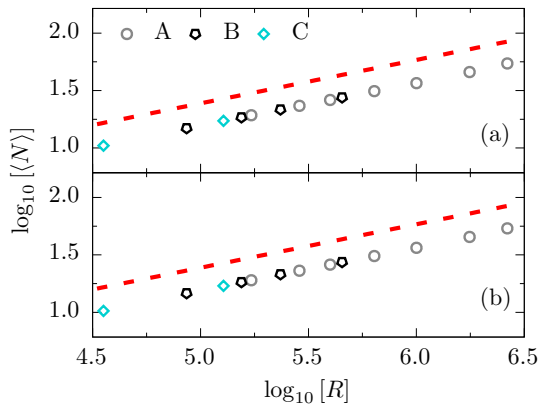}
\caption{Scaling of the average number of hops at low temperatures for
the time intervals (a) T50 and (b) T100. Dashed lines have slope $0.38$.
The parameter sets are presented in Table \ref{energias}.}
\label{smallR}
\end{figure}

In Fig. \ref{smallR}(a), we plot $\langle N\rangle$ as a function of $R$ in T50,
considering three parameter sets and temperatures that give $R\epsilon\leq 0.2$.
In Fig. \ref{smallR}(b), we plot the data obtained in T100 with the same conditions.
Both plots suggest a power-law relation
\begin{equation}
\langle N\rangle \sim R^a ,
\label{Nlow}
\end{equation}
and linear fits of their data give $a=0.38\pm 0.01$.

\subsection{Average number of hops at intermediate or high temperatures}
\label{averagehigh}

High temperature conditions are typically the cases with
$R>{10}^6$ and $\epsilon >{10}^{-2}$, in which simulations indicate that
$\langle N\rangle$ depends on both $R$ and $\epsilon$.
Since the surface mobility is high, the hopping of the
adatom is expected to be limited by the time $1/F=R/D_0$ necessary for deposition of
a new layer (which buries the previous layer), instead of being limited by attachment to NNs.
Indeed, our simulations with approximately constant
$\epsilon$ indicate that $\langle N\rangle$ is approximately proportional to $R$.
This leads to the proposal
\begin{equation}
\langle N\rangle \sim R \epsilon^b
\label{Nhigh}
\end{equation}
at sufficiently high temperatures, where $b$ is an exponent to be determined numerically.

However, it is very difficult to fit Eq. (\ref{Nhigh}) because
the number of data points with large $R$ is small (the simulations are very time consuming)
and it is difficult to anticipate a reliable criterion for the high temperature limit.
As an alternative, we look for a scaling relation that incorporates low and high temperature
behaviors and the crossover in intermediate temperatures, in which a larger number of data points
may be available.

A scaling relation of general validity for $\langle N\rangle$ has to be consistent with
Eqs. (\ref{Nlow}) and (\ref{Nhigh}).
The consistency with the latter suggests the form
\begin{equation}
\langle N\rangle \sim R \epsilon^b f\left( x\right) \qquad , \qquad x \equiv R\epsilon^c ,
\label{Nscaling}
\end{equation}
where $f$ is a function that converges to a constant as $x\to\infty$
(i.e. very high temperatures), and $c$ is another exponent to be determined numerically.
The consistency with Eq. (\ref{Nlow}) leads to the exponent relation
\begin{equation}
c = \frac{b}{1-a} .
\label{defc}
\end{equation}
Since the value $a=0.38$ is known, Eq. (\ref{defc}) implies that the values of $b$ and $c$
are not independent, and we have to search only for the numerical estimate of one of them.

\begin{figure}[ht]
\center
\includegraphics [scale=1.0]{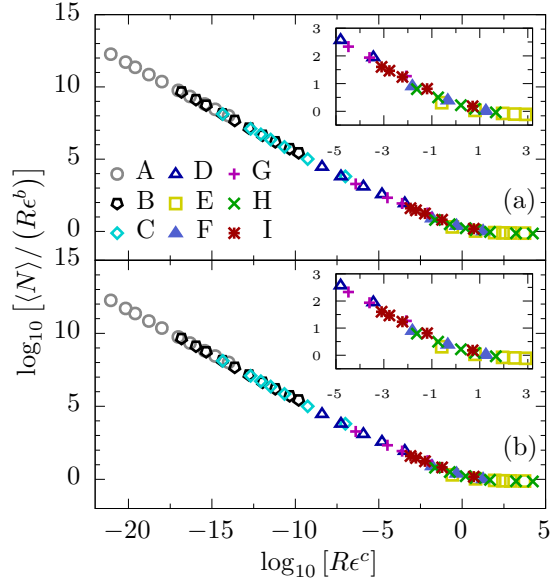}
\caption{Scaling plot of the average number of adatom hops using $b=1.85$ and $c=3.0$ for
the time intervals (a) T50 and (b) T100.
The uncertainties are smaller than the size of the data points.
The parameter sets are in Table \ref{energias}.
The insets show magnified zooms near the crossover regions.}
\label{colapso}
\end{figure}

Following Eq. (\ref{Nscaling}), we plotted $\langle N\rangle /\left( R \epsilon^b \right)$
as a function of the scaling variable $x$ for several values of the
exponent $b$; in each case, $c$ obtained from Eq. (\ref{defc}).
Good collapse of our data is obtained for $1.7\leq b\leq 2$, which corresponds to
$c=3.0\pm0.3$.
Figs. \ref{colapso}(a) and \ref{colapso}(b) show scaling plots
using the central estimates of these exponents ($b=1.85$, $c=3.0$) in T50 and T100, respectively.
Those plots span more than $20$ decades of the abscissa $x$ and more than $10$ decades
of $\langle N\rangle /\left( R \epsilon^b \right)$.
The insets of Figs. \ref{colapso}(a) and \ref{colapso}(b) show magnified zooms of some regions
of the main plots, which permits the observation of uncertainties in the data.

In Figs. \ref{colapso}(a) and \ref{colapso}(b), the data points
with $R\epsilon^3\geq {10}^2$ seem to converge to a constant value.
Thus, this relation numerically defines the high temperature regime.
The small number of data points in those conditions explains the difficulty to fit
Eq. (\ref{Nhigh}).
To observe the asymptotic behavior with constant
$\langle N\rangle /\left( R \epsilon^b \right)$, several simulations
in much larger values of $R\epsilon^3$ would be necessary.

The results presented in this section and in the previous one also show that there
is no significant difference in the values of $\langle N\rangle$ obtained in the time
intervals T50 and T100, and, within error bars, no difference in scaling exponents.

\subsection{Distributions of  numbers of hops}
\label{distributions}

When the distributions have universal shapes, they can be written in the form
\begin{equation}
P\left( N\right) = \frac{1}{\langle N\rangle} g\left( \frac{N}{\langle N \rangle} \right) ,
\label{scalingP}
\end{equation}
where $g$ is a scaling function.
For instance, similar form was already shown to fit roughness
distributions of growth models \citep{antal2001}.
Here we check this ansatz separately for low and for intermediate to high temperatures.

\begin{figure}[!ht]
\center
\includegraphics [scale=0.95]{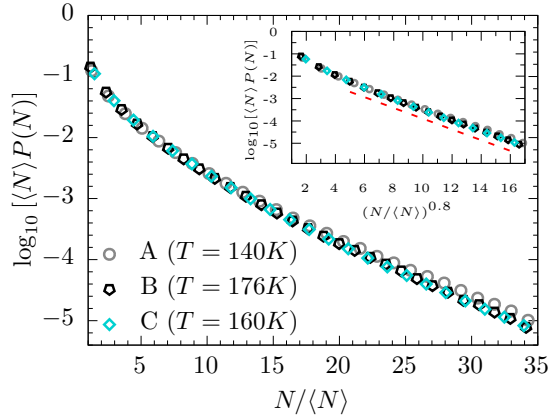}
\caption{Scaled distributions in the low temperature regime, for the parameter sets and temperatures
indicated in the plot.
The inset shows the same results with the variable
${\left( N/\langle N \rangle\right)}^{0.8}$ in the abscissa (the dashed line is a guide to the eye).}
\label{distrlow}
\end{figure}

Fig. \ref{distrlow} shows the scaled distributions for three parameter sets
in the low temperature regime in the interval T50.
The upward curvatures of the plots suggest stretched exponential tails,
in the form $\exp{\left[ -{\left( N/\langle N \rangle\right)}^{\gamma}\right]}$ with $\gamma<1$.
This is confirmed in the inset of Fig. \ref{distrlow} using $\gamma=0.8$;
other choices of this exponent also lead to good data fits and an estimate
$\gamma=0.80\pm 0.07$.

\begin{figure}[!ht]
\center
\includegraphics [scale=0.9]{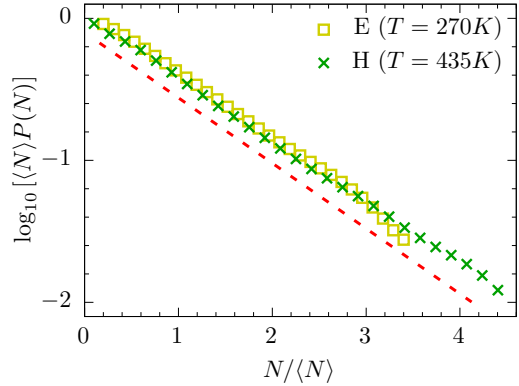}
\caption{Scaled distributions for two parameter sets and temperatures, in the high temperature regime.
The dashed line is a guide to the eye.}
\label{distrhigh}
\end{figure}

Fig. \ref{distrhigh} shows two scaled distributions in the high temperature regime
defined in Sec. \ref{averagehigh}, i.e. $R\epsilon^3\gg 1$, in the interval T50.
In these cases, the tails can be fit as simple exponential decays. We stress that there is 
no difference between the results presented for T50 and T100.

\subsection{Results for the LM model}
\label{LMresults}

\begin{figure}[!ht]
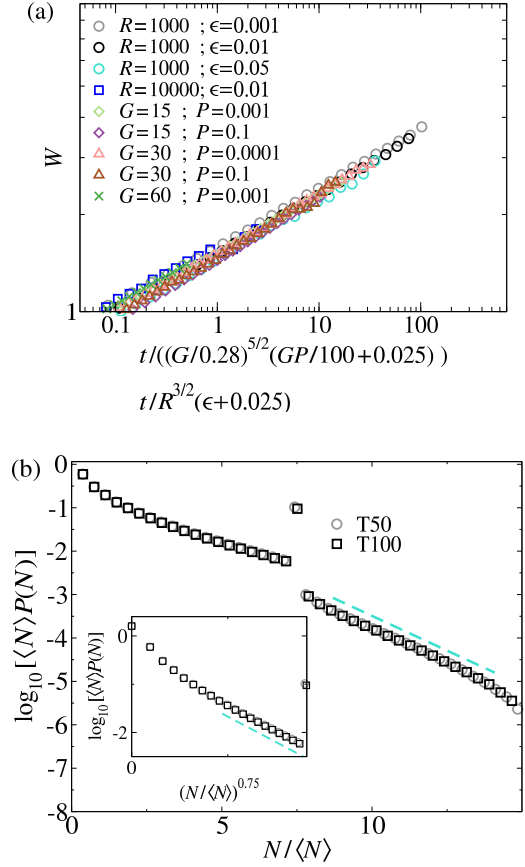

\center
\includegraphics [height=5.5cm]{Colapso_w_ICV1eCV_nova.eps}\\
\vspace{0.5cm}
\includegraphics [height=5.5cm]{dist_G50P0p01.eps}
\caption{(a) Roughness as a function of scaled times in the CV and the LM model for the parameters
indicated in the plot.
(b) Distributions of the number of executed hops in the LM model for
$G=60$ and $\epsilon=0.01$.
The inset shows a rescaling of the data for $N<G$
with ${\left( N/\langle N \rangle\right)}^{0.75}$ in the abscissa.
The dashed lines in the main plot and in the inset are guides to the eye.
}
\label{distrLM}
\end{figure}

In Fig. \ref{distrLM}(a), we show the roughness evolution of the LM and CV models with different
parameters and a suitable scaling of the growth time with their parameters.
The rescaling of the CV data is similar to that of Ref. \protect\cite{cv2015}; the
rescaling by $G^{5/2}$ in the LM model is based on the results of Ref. \protect\cite{CDLM},
and the product $GP$ is incorporated with a role similar to that of $\epsilon$.
The reasonable data collapse in Fig. \ref{distrLM}(a), with slopes near $0.2$ for all parameter sets,
confirms that both models have the same roughness scaling.
This rougheness scaling is described by the Villain-Lai-Das Sarma
(VLDS) equation in the hydrodynamic limit \citep{villain,laidassarma}.

Fig. \ref{distrLM}(b) shows the distributions of numbers of executed hops in the LM model in
the time intervals T50 and T100, for $G=60$ and $P=0.01$; no rescale was used in these plots.
In these films, the number of detachments from lateral NNs per atom was near $0.1$, which
means that such detachments were not frequent.
The distributions have peaks at $N=G$, which originate from adatoms that executed all the
initial $G$ hops; most of them probably moved on terraces all the time.
For $N<G$, the distributions seem to have stretched exponential forms, which is confirmed in
the inset of Fig. \ref{distrLM}(b) by the rescaling with the abscissa ${\left( N/\langle N \rangle\right)}^{0.75}$.
The main contribution to this part of the distribution is from adatoms that stopped moving after the
first attachment to a NN.
For $N>G$, Fig. \ref{distrLM}(b) suggests a simple exponential decay;
this part of the distribution corresponds to adatoms that detached at least once from lateral NNs.

Since these distributions reproduce features of the low and high temperature regimes of the CV model,
we understand that those features are not particular of that model and that they are related
to the processes of lateral attachment and detachment.

\section{Scaling approach}
\label{discussion}

\subsection{Trapping of random walkers and the low temperature regime}
\label{trapping}

The results for the LM model indicate that the adatoms that execute
$N<G$ hops are those that move on terraces and permanently stick when they
reach the terrace borders (terrace steps), at the same height or hopping to a lower height.
Thus, those borders act almost as perfect traps.
We propose that the shape of the distributions are related to this trapping phenomenon.
Since the distributions in the CV model at low temperatures have similar upward curvature,
are fit by similar stretched exponentials, and the CV roughening is the same as
the LM model, we expect that the same trapping phenomenon occurs.
The main difference is that, in the CV model, the landscape in which the adatom moves
(including terrace borders) dynamically evolves, while in the LM model it is static.

For these reasons, here we search for possible relation with a much simpler problem:
the trapping of a random walker on a surface with a random distribution
of static traps with density $\rho$.
Approximate \citep{havlin} or exact \citep{donsker1979} solutions of this problem show that,
at very long times, the probability that the walker survives without being trapped is
an exponential of the variable ${\rho D_W t}^{1/2}$,
where $D_W$ is the walker diffusion coefficient.
However, previous simulations in large system sizes with trap density $0.5$ could not
fit this theoretical prediction; instead, they showed stretched exponential
distributions with exponents larger than $1/2$ \citep{bunde1997}.

\begin{figure}[ht]
\center
\includegraphics [height=5.5cm]{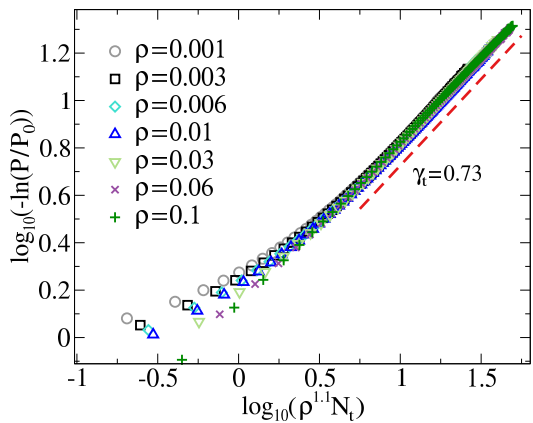}
\caption{Scaled distributions of the number of executed hops in the random trap model with
three different densities.
$P_1$ is the probability of survival after a single hop.}
\label{disttrap}
\end{figure}

Here we performed our own simulations with ${10}^9$ random walks in very large square lattices
and trap densities from $0.001$ to $0.1$.
We measured the distributions
$P_t\left( N_t\right)$ of the number of executed hops $N_t$, where the subindex $t$ refers
to the trapping problem.
In Fig. \ref{disttrap}, we show an excellent data collapse of the distributions after
a suitable rescaling of $\rho$ and $N_t$.
Accounting for the different rescalings
that also produce a good collapse of those distributions, we obtain
\begin{equation}
P_t\left( N_t\right) \sim \exp{\left[ -{\left(N_t/\langle N_t\rangle\right)}^{\gamma_t}\right]} \qquad , \qquad
\gamma_t = 0.75\pm 0.05 ,
\label{scalingPt}
\end{equation}
where
\begin{equation}
\langle N_t\rangle \sim \rho^{-\lambda}  \qquad , \qquad \lambda =1.10\pm 0.02 .
\label{scalingNt}
\end{equation}
This is an effective (not asymptotic) scaling, but it is a good approximation
from $N_t\sim 1$ to $N_t\sim {10}^4$.

The exponent $\gamma$ obtained in the distributions of the CV model at low temperatures is
very close to $\gamma_t$ in the trapping model.
Although the distribution of the ``trapping terrace borders" in the CV model is spatially
correlated, we believe that their fluctuations reduce the effect of these correlations
from the point of view of the atoms that diffuse on the terraces.

To provide additional support to this interpretation, we analyze the effect of the trap density.
Simulations of the CV model with $\epsilon=0$ (which is representative of low temperatures)
show the formation of terraces whose average area scales as $A\sim R^{0.6}$ \citep{CDLM}.
Assuming that these terraces have relatively compact borders, the density of surface steps $\rho_s$
is expected to scale approximately as the perimeter-to-area ratio, i.e. $\rho_s\sim R^{-0.3}$.
The average number of hops in the CV model, which is given in Eq. (\ref{Nlow}),
can be written in terms of this density as
$\langle N\rangle\sim \rho_s^{-0.38/0.3} = \rho_s^{-1.3}$.
The exponent $1.3$ in this relation is not very distant from the effective exponent $1.1$ of
the trapping model [Eq. (\ref{scalingNt})].
The discrepancy may be a consequence of not accounting for the disorder in the terrace
borders, which leads to a different perimeter-to-area ratio, and for the approximation underlying
the assumption of randomness in the trap distribution.

In the above reasoning, the kinetic roughening of the CV model plays a role in the scaling of the
average area because the exponent $0.6$ stands for the ratio $2/z$,
where $z$ is the dynamical exponent of the VLDS class \citep{villain,laidassarma}.
However, $R$ is a non-universal parameter in that context.
Consequently, the above results establish a connection between universal and non-universal
features of the CV model and the two-dimensional random trapping problem.

\subsection{Smooth surfaces in intermediate to high temperatures}
\label{smooth}

In the scaling plots of Figs. \ref{colapso}(a)-(b), deviations from the linear decay
(low temperature behavior) are observed for $R\epsilon^3\gtrsim {10}^{-3}$.
This is the region analyzed here, which we generically term intermediate to high temperatures.

\begin{figure}[ht]
\center
\includegraphics [scale=0.85]{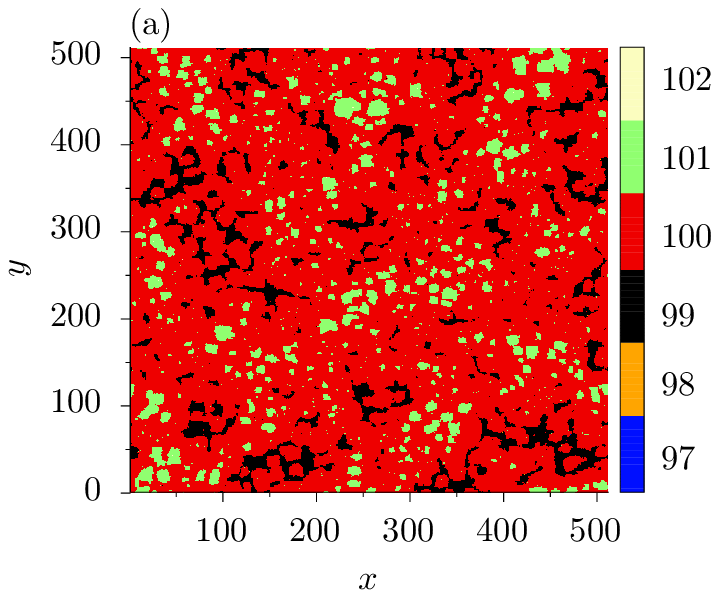}
\includegraphics [scale=0.85]{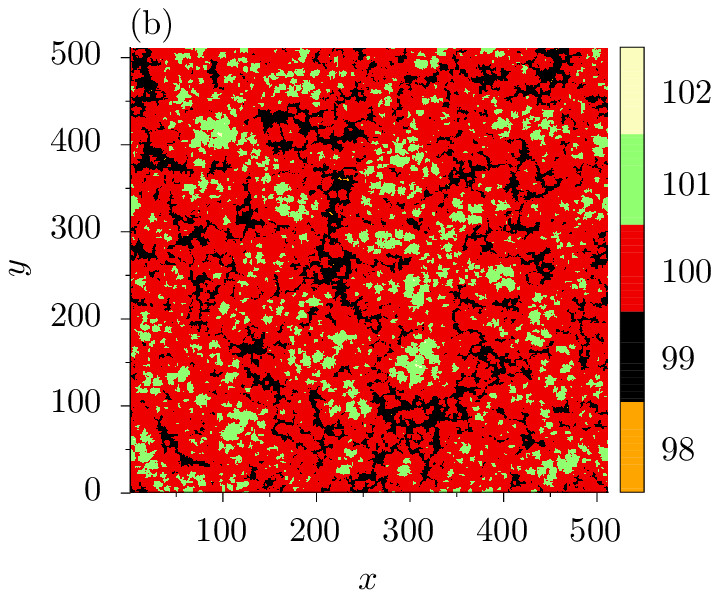}
\includegraphics [scale=0.85]{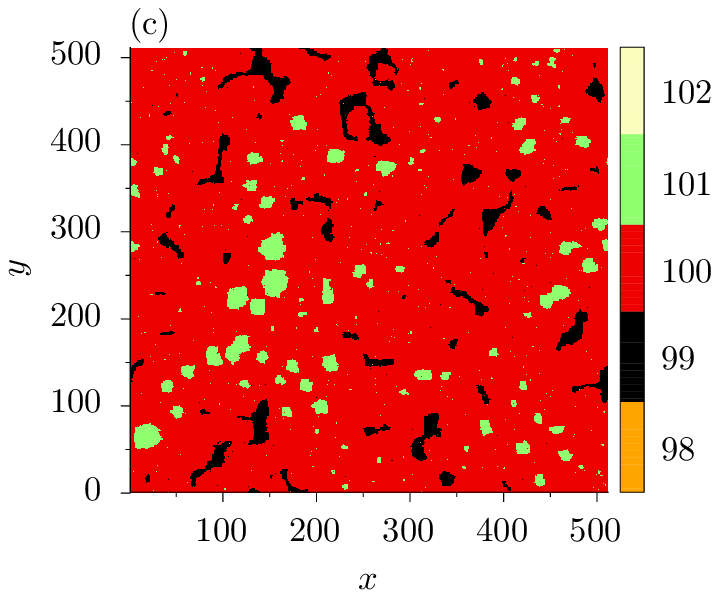}
\caption{Top view of films grown with the parameter sets (a) H ($T=340$K), (b) I ($T=560$K),
and (c) F ($T=253$K).}
\label{surfaces}
\end{figure}

The surface roughness in this regime is always smaller than $1$ in T100 and T50, which means that
the local height is typically equal to the average or fluctuates one unit above or below
the average.
Figs. \ref{surfaces}(a)-(c) show top views of three surfaces after deposition of
$100$ layers, which confirm that they are very smooth.

At very high temperatures, the average number of hops
scales as Eq. (\ref{Nhigh}) with $b$ close to $2$.
Observation of Fig. \ref{surfaces}(c) shows that the corresponding surfaces have very large
terraces.
This morphology resembles that of islands in high temperature submonolayer growth,
i.e. with an incomplete layer being formed on a flat surface of the same material.

For intermediate coverages ($0.05{\text -}0.30$),
the submonolayers produced by the CV model in those conditions
have large islands with a rounded shape, which are surrounded by a free adatom gas
with density $\rho_f\sim \epsilon^2$ \citep{submonorev}.
These free adatoms have no NN and can rapidly move in the region between the islands;
see e.g. illustrations in Ref. \protect\cite{submonorev}.
That value of the density is related to a balance between the random attachment of the free adatoms
to the islands and a preferential detachment at kink sites ($n=2$) of the island borders.

In those high temperature submonolayers or in the high temperature films,
the adatoms at island borders are more tightly bound, so the average
number of hops in a given time interval is dominated by those free adatoms.
In a time interval ${\delta t}_F=1/F$, a new layer of adatoms covers the relatively smooth surface,
so this is the time in which a given set of free adatoms is expected to move
with coefficient $D_0$.
Thus, the average number of hops is $\langle N\rangle\sim \rho_f D_0{\delta t}_F=R\epsilon^2$,
which is in good agreement with our numerical estimate for the exponent $b$.

These arguments also suggest that the adatom diffusion is limited by the deposition of new atomic layers.
Since there is a random atomic flux, an adatom that executed a given number of hops has a
constant probability of not being able to execute an additional hop.
This explains the simple exponential decay of $P\left( N\right)$ in intermediate to high temperatures.

Our numerical results also predict that the crossover from the low to the high temperature regime
occurs when $R \epsilon^3\sim 1$.
In the characteristic time ${\delta t}_F$, we have $R \epsilon^3=D_0 \epsilon^3{\delta t}_F$, which
is the probability of an adatom to detach from a straight terrace step
where it has $3$ lateral NNs.
Consequently, we interpret the crossover as the condition in which detachment of adatoms from
straight terrace steps begins to be relevant
(of course the detachment is much more frequent from angular parts of those steps, in
which the number of NNs is $1$ or $2$).

As a final note, the above mentioned results of Ref. \protect\cite{submonorev}
account for island growth control by attachment and detachment of adatoms from their borders,
even in the conditions of low flux.
The dominant role of the attachment/detachment kinetics is also observed in studies of
post-deposition coarsening in the CV model,
in which other processes such as island diffusion and coalescence may be relevant
only at early times after the deposition has stopped \citep{LAM1999,Shi2007}.

\section{Conclusion}
\label{conclusion}

We studied the scaling properties of the distribution of the number of hops of adatoms
at the surface of growing films in the Clarke-Vvedensky model in a simple cubic lattice.
Various sets of energy parameters and temperatures were considered, with no barrier for crossing
step edges.
We distinguish low and high temperature regimes, which are respectively characterized by
irreversible adatom aggregation to a single lateral neighbor and by possible adatom detachment from
terrace borders with up to three lateral neighbors.

At low temperatures, the scaled distributions show a stretched exponential decay and the average
number of hops scales with the diffusion-to-deposition ratio $R$, but does not depend on the
detachment probability $\epsilon$ from lateral neighbors.
The stretched exponential is interpreted in terms of a trapping of the diffusing adatoms by
the borders of the terraces in which they begin to move after adsorption.
Simulations of a two-dimensional trapping model show a similar stretched exponential decay of
the survival probability (which is an effective scaling valid up to $\sim {10}^4$ hops).
They also show that the effect of the trap density is
similar to the effect of the density of terrace steps on the average number of hops of the CV adatoms.
At high temperatures, the scaled distributions have a simple exponential decay,
which is related to the randomness of the flux that eventually covers the previous atomic layer.
The average number of hops is dominated by freely moving atoms in a relatively smooth surface,
and the scaling with $R$ and $\epsilon$ is explained by results of submonolayer growth.

These results show that kinetic roughening theory and non-equilibrium reaction-diffusion problems,
such as the trapping models, may be useful to describe the statistics of adatom hops in a
film growth model.
This is particularly important when the film landscape is responsible for constraining the adatom
diffusion.
Here, this was the case of low temperatures, but in more complex growth models with different
energy barriers, the film surfaces may be rough or have patterns at higher temperatures.
Our results may also be important for the design of novel models of thin film deposition
that mimic the CV model or its variants, and which have been studied by several authors in
recent years \citep{limkumnerd2014,disrattakit2016,tosousareis2018,jinAIPAdv2018,martynec2019,
pereira2019,blel2019}.

From the distributions of the numbers of hops and from the type of diffusion executed by the
adatoms (normal or anomalous), it is possible to obtain distributions of diffusion lengths
and average values of those lengths.
With the advance in imaging techniques, some of these results may be tested
experimentally, despite the difficulties that are expected to monitor the diffusion
of an atom while a film of the same material grows.


\begin{acknowledgments}

FDAAR is supported by Brazilian agencies CAPES (88881.068506/2014-01),
CNPq (305391/2018-6), and FAPERJ (E-26/202.881/2018).
ISSC is supported by FAPERJ (E-26/202.356/2018).
TAdA is supported from CNPq (308343/2017-4).
EEML is supported from CAPES.

\end{acknowledgments}



%

\end{document}